%% file: main.tex
\documentclass[conference]{IEEEtran}
\IEEEoverridecommandlockouts

\usepackage{graphicx}
\usepackage{xcolor}
\usepackage{multirow}
\usepackage{booktabs}
\usepackage{amsmath}
\usepackage{amssymb}
\usepackage{cite}
\usepackage{cuted}
\newcommand{\linebreakand}{%
  \end{@IEEEauthorhalign}
  \hfill\mbox{}\par
  \mbox{}\hfill\begin{@IEEEauthorhalign}
}    
\usepackage{bbm}
\usepackage{algorithm} 
\usepackage{algpseudocode}
\usepackage{caption}
\usepackage{steinmetz}
\usepackage{subcaption}
\usepackage{amsthm}
\usepackage[english]{babel}


\newcounter{relctr} 
\everydisplay\expandafter{\the\everydisplay\setcounter{relctr}{0}} 
\usepackage[figurename=Fig.]{caption}
\newcommand{\norm}[1]{\left\lVert#1\right\rVert}
\usepackage{xpatch}

\newtheoremstyle{remarkstyle}%
  {}
  {}
  {\itshape}
  {}
  {\itshape}
  {.}
  {.5em}
  {}

\theoremstyle{remarkstyle}

\AtBeginDocument{} 

\ifCLASSINFOpdf
\else
\fi
\usepackage{acro}
\input{acronyms}

\begin{document}

\title{Beam Selection for Delay-Doppler Visibility in Multi-Target MIMO-OFDM Sensing \vspace{-2mm}\\
\thanks{This work is supported by the Research Council of Finland (Grants 362782 (ECO-LITE), 369116 (6G Flagship)).}

\author{
{ Amirhossein~Azarbahram and Onel~L.~A.~L\'opez}
\vspace{1mm}
\\

\small Centre for Wireless Communications, University of Oulu, Finland. Emails: \{amirhossein.azarbahram, onel.alcarazlopez\}@oulu.fi}
}

\maketitle

\begin{abstract}
This paper studies leakage-aware beam selection for multi-target multiple-input multiple-output (MIMO)-orthogonal frequency-division multiplexing (OFDM) sensing. We focus on ensuring that each hypothesized target remains detectable at its own delay-Doppler (DD) bin despite leakage from other targets. For this, we derive a visibility metric that separates the desired focused power from pairwise leakage caused by the finite-grid OFDM ambiguity kernel, receive angular coupling, and transmit codebook gains. Then, we formulate a max-min beam-selection and power-allocation problem in which the sensing signal over the fixed OFDM time-frequency grid is transmitted by a limited number of selected beams. The resulting problem is solved via a bisection-based mixed-integer successive convex approximation framework. A time-sharing formulation is also presented as a globally solvable benchmark. Numerical results show that the proposed pairwise leakage-aware design achieves higher worst-target visibility than the time-sharing, global leakage-suppression, illumination-based, and random beam-selection baselines, especially when targets are close in angle or DD.
\end{abstract}


\section{Introduction}

Future wireless communication systems are expected to embed sensing capabilities, giving rise to the \ac{isac} paradigm \cite{3gpp_tr_22870_v2000}. 
Since cellular \ac{isac} is expected to reuse communication infrastructure, spectrum, and frame structures, many practical designs are constrained by the underlying communication waveform. In this context, \ac{ofdm} is especially relevant, since \ac{cp}-\ac{ofdm} is the baseline downlink waveform in \ac{nr}. From a sensing perspective, the \ac{ofdm} time-frequency grid is also useful because target echoes can be processed across subcarriers and symbols to obtain \ac{dd} information, where delay is related to range and Doppler to radial motion. In a \ac{mimo}-\ac{ofdm} setting, the antenna arrays add a spatial dimension allowing the transmit beams to determine how strongly different target directions are illuminated, while receive combining provides angular selectivity among their echoes~\cite{TS38300,correlation_radar}.


Reusing a communication \ac{ofdm} grid for sensing shifts transmit design toward controlling target illumination under a fixed \ac{dd} structure. After \ac{dd} focusing, each target must appear around its own bin, while finite duration and bandwidth lead to a non-impulsive finite-grid response that is determined by the occupied time-frequency samples appearing as a \ac{dd} ambiguity kernel~\cite{correlation_radar}. Thus, one target can leak into another target's focused bin, especially when their separation is small or fractional with respect to the grid, or when a strong target masks a weaker one, while the masking severity depends jointly on the ambiguity kernel, receive angular coupling, and transmit beam gains toward targets~\cite{mimo_ofdm_tut,waveform_comradar}.

Leakage effects have been addressed via global sidelobe suppression formulations under communication and sensing constraints, e.g.,~\cite{SLP_sidelobe,liu2025rangeangle,moveble_isl}. For \ac{mimo}-\ac{ofdm} \ac{isac}, \cite{SLP_sidelobe} designs the transmit signal using symbol-level precoding to minimize the range-Doppler \ac{isl},
while digital beamforming suppresses the range-angle \ac{isl} ratio in \cite{liu2025rangeangle}.
The work in \cite{moveble_isl} expands the design space by jointly optimizing transmit beamforming and movable-antenna positions for range-Doppler \ac{isl} suppression. More generally, beamforming-oriented \ac{isac} designs optimize beam patterns, sensing gains, estimation metrics, or communication-sensing tradeoffs using continuous beamforming variables~\cite{opt_BF_ISAC,isac_txrx_bf}. 

The above approaches rely on design freedoms not available when sensing reuses a fixed communication waveform and/or when the transmitter uses a finite beam codebook, as the sensing layer cannot freely reshape the \ac{ofdm} grid or synthesize arbitrary continuous beamformers. The remaining design freedom is to select several probing beams and allocate sensing resources to them. This is within the scope of beamspace and codebook-based \ac{isac} methods, including multi-beam transmission and beam sweeping for mmWave massive \ac{mimo} \ac{isac}, e.g., \cite{beamsweep_isac,multibeamisac_mag}. However, these works mainly target beam coverage, joint communication-sensing illumination, or sweeping-overhead reduction, rather than the target visibility in \ac{dd} domain. Motivated by this gap, we consider a multi-target \ac{mimo}-\ac{ofdm} \ac{isac} system, where the transmitter probes the scene using a finite beam codebook in a given OFDM grid by selecting a limited number of codebook beams and allocating their powers, aiming to keep all hypothesized targets visible at their own \ac{dd} bins. 

The main contributions are fourfold. \emph{\textbf{First,}} we derive a target-bin visibility metric and use it to formulate a finite-codebook beam-selection and power allocation problem to maximize the worst target visibility in the \ac{dd} domain. \emph{\textbf{Second,}} we develop a bisection-\ac{mi}-\ac{sca} algorithm for the resulting mixed-integer fractional problem, where each fixed-visibility feasibility test is handled through a conservative convex approximation. \emph{\textbf{Third,}} we propose a time-sharing formulation, solved optimally by a bisection \ac{mi} \ac{lp} framework, as a structured initialization for the simultaneous transmission design. \emph{\textbf{Fourth,}} we show numerically that the proposed leakage-aware beam selection improves worst-target visibility compared to the time-sharing, global leakage-suppression, and illumination-based baselines, especially when targets are close in angle or \ac{dd} domains.


\textbf{\textit{Notation:}} Bold lowercase letters denote vectors. The operators $(\cdot)^T$, $(\cdot)^H$, and $(\cdot)^*$ denote transpose, Hermitian transpose, and complex conjugation, respectively. The notation $\norm{\cdot}$ denotes the $\ell_2$-norm operator, while $\mathbb E\{\cdot\}$ denotes statistical expectation. $\mathcal{CN}(\mathbf 0,\sigma^2\mathbf I)$ denotes a circularly symmetric complex Gaussian random vector with covariance $\sigma^2\mathbf I$.

\section{Sensing System and Signal Model}\label{sec:sysmod}

\begin{figure}
    \centering    \includegraphics[width=0.97\linewidth]{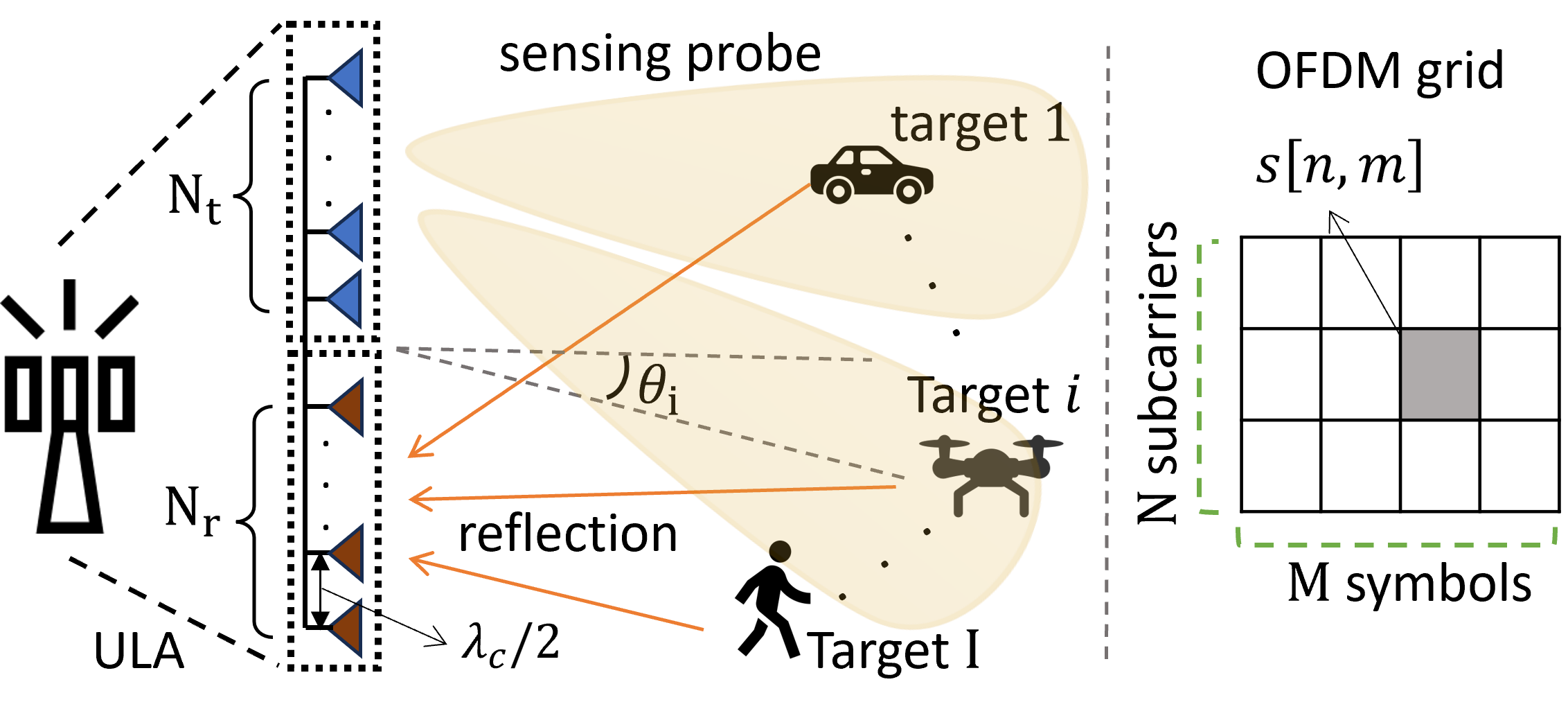}
    \vspace{-1mm}
    \caption{Multi-target monostatic MIMO-OFDM sensing model.}
\label{fig:sysmod}
\vspace{-5mm}
\end{figure}

We consider a monostatic \ac{ofdm} \ac{ula} transceiver with $N_t$ transmit and $N_r$ receive antennas. The carrier frequency is $f_c$, and the wavelength is $\lambda_c=c/f_c$, where $c$ is the speed of light. The \ac{ofdm} frame contains $N$ subcarriers and $M$ OFDM symbols. The subcarrier spacing is $\Delta f$, and the OFDM symbol duration including \ac{cp} is $T_{\rm sym}$. The transmitted \ac{ofdm} grid is denoted by $s[n,m]$, where $n=0,\ldots,N-1$ and $m=0,\ldots,M-1$. The system reflects a fixed-waveform \ac{isac} sensing setting, where the probing signal uses an \ac{ofdm} grid compatible with the underlying communication waveform and frame structure. The system model is illustrated in Fig.~\ref{fig:sysmod}.

The transmitter uses a finite beam codebook $\mathbf W=[\mathbf w_1,\ldots,\mathbf w_B]$, with $\norm{\mathbf w_b}=1$, $\forall b$, $\mathbf W^{H}\mathbf W=\mathbf I_B$, and non-negative beam coefficient $P_b\geq 0$. The transmitted signal is
\begin{equation}
\mathbf x[n,m]
=
\sum_{b=1}^{B}
\sqrt{P_b}\mathbf w_b s[n,m]
=
\mathbf W\boldsymbol\alpha s[n,m],
\label{eq:sim_tx_signal}
\end{equation}
where
$\boldsymbol\alpha=[\sqrt{P_1},\ldots,\sqrt{P_B}]^T$.
%
%
Let $\mathcal{T}=\{1,\ldots,I\}$ be the set of targets, with target $i$ described by $(\theta_i,\ell_i,\nu_i,\xi_i)$, where $\theta_i$ is the angle of target $i$ with respect to the array broadside, $\ell_i=N\Delta ft_i$ is the normalized delay-bin coordinate, $\nu_i=M T_{\rm sym} f_{D,i}$ is the normalized Doppler-bin coordinate, and $f_{D,i}$ is its monostatic Doppler frequency. For radial velocity $v_i$, the Doppler frequency is given by $f_{D,i}
={2v_i f_c}/{c}$. The coefficient $\xi_i$ absorbs all environmental and target effects, including two-way propagation, path loss, and radar cross-section. The set of target hypotheses is known from a preceding coarse sensing stage, a tracker, or an external sensing update. This is consistent with cellular sensing-service scenarios where the network supports object detection/tracking. Moreover, delay, Doppler, angle, and signal strength are sensing quantities obtainable from scattered and reflected \ac{nr} signals~\cite{3gpp_ts_22137}.



Assuming transceiver self-interference cancellation and using~\eqref{eq:sim_tx_signal}, the received signal is given by
\begin{align}
\mathbf y[n,m]
&=\sum_{i=1}^{I}
\xi_i\,\mathbf a_r(\theta_i)\mathbf a_t^H(\theta_i)\mathbf W\boldsymbol\alpha\,s[n,m]
\nonumber\\
&\quad \times e^{-j2\pi n\ell_i/N}e^{j2\pi m\nu_i/M}
+\mathbf z[n,m],
\label{eq:rx}
\end{align}
where $\mathbf a_t(\theta_i)
=[1,e^{\jmath\pi\sin\theta_i},\ldots,
e^{\jmath\pi(N_t-1)\sin\theta_i}]^T$ and
$\mathbf a_r(\theta_i)
=[1,e^{\jmath\pi\sin\theta_i},\ldots,
e^{\jmath\pi(N_r-1)\sin\theta_i}]^T$
are the transmit and receive steering vectors for $\lambda_c/2$-spaced \ac{ula}, respectively, and $\mathbf z[n,m]\sim\mathcal{CN}(\mathbf 0,\sigma_z^2\mathbf I)$ is additive white Gaussian noise. We assume that the \ac{cp} is long enough to avoid inter-symbol interference and that Doppler-induced inter-carrier interference is negligible, as in~\cite{SLP_sidelobe,moveble_isl}.


We compensate for $s[n,m]$ before sensing and receive processing. Defining $E_s \triangleq \sum_{n=0}^{N-1}\sum_{m=0}^{M-1}|s[n,m]|^2/(NM)$, $\tilde{\mathbf y}[n,m]={s^*[n,m]}\mathbf y[n,m]/{E_s}$, and substituting~\eqref{eq:rx} into this compensated grid shows that each target contribution is weighted by the normalized grid-energy, i.e., $w[n,m]\triangleq|s[n,m]|^2/E_s$. Without loss of generality, we normalize the \ac{ofdm} grid such that $E_s=1$.

For a trial processing bin $(\ell,\nu)$, \ac{dd} focusing applies the phase compensation $e^{j2\pi n\ell/N}e^{-j2\pi m\nu/M}$ and averages over the $NM$ grid samples. Hence, a target located at $(\ell_j,\nu_j)$ contributes to the focused bin through the offsets $\Delta\ell=\ell-\ell_j$ and $\Delta\nu=\nu_j-\nu$. This gives the fixed ambiguity kernel induced by the occupied \ac{ofdm} grid as~\cite{correlation_radar}
\begin{equation}
A(\Delta\ell,\Delta\nu)
=\frac{1}{NM}\sum_{n=0}^{N-1}\sum_{m=0}^{M-1} w[n,m]
 e^{j2\pi (\Delta\ell\frac{n}{N} + \Delta\nu\frac{m}{M})}.
\label{eq:kernel}
\end{equation}
If a target has fractional \ac{dd} coordinates, the finite exponential sums no longer cancel at neighboring grid bins, spreading the target response across the ambiguity kernel, creating leakage.

\section{Target Visibility and Problem Formulation}\label{sec:metricform}

For target $i$, the compensated received signal is focused at $(\ell_i,\nu_i)$ and then combined toward $\theta_i$ using the normalized receive combiner $\mathbf a_r^H(\theta_i)/\sqrt{N_r}$. The resulting observation contains the desired response of target $i$, the leakage contributions from the other targets, and the focused noise. Hence, it can be written as $\hat r_i
=
\sum_{j\in\mathcal T} r_{j\to i}
+
\tilde z_i$, where $r_{j\to i}$ denotes the contribution of target $j$ to target $i$'s focused bin, and $\tilde z_i$ is the noise after symbol compensation, \ac{dd} focusing, and receive combining. Since the receive signal processing operations are linear, $\tilde z_i$ remains zero-mean complex Gaussian. Considering the normalization in~\eqref{eq:kernel} and the normalized receive combiner, its variance is $\eta = \mathbb E\{|\tilde z_i|^2\}
={\sigma_z^2}/({NM E_s})$.

Using the received signal model, the contribution of target $j$ is given by
\begin{align}
r_{j\to i}
\!=\!
\xi_j
\frac{\mathbf a_r^H(\theta_i)\mathbf a_r(\theta_j)}{\sqrt{N_r}}
\mathbf a_t^H(\theta_j)\mathbf W\boldsymbol\alpha
A(\ell_i\!-\!\ell_j,\nu_j\!-\!\nu_i),
\label{eq:pair_contrib}
\end{align}
which separates the three coupling mechanisms that determine how target $j$ appears in target $i$'s focused bin: i) the factor $\mathbf a_r^H(\theta_i)\mathbf a_r(\theta_j)/\sqrt{N_r}$ captures receive-side angular coupling; ii) $\mathbf a_t^H(\theta_j)\mathbf W\boldsymbol\alpha$ is the transmit gain of the composite spatial beam toward target $j$; and iii) $A(\ell_i-\ell_j,\nu_j-\nu_i)$ is the \ac{dd} coupling induced by the finite \ac{ofdm} ambiguity kernel. For $j=i$, we have $A(0,0)=1$, and the term in~\eqref{eq:pair_contrib} is the desired focused response, while for $j\neq i$, the same expression represents leakage from target $j$ into target $i$'s bin.

For notation simplicity, define
$\mathbf g_j=\mathbf W^H\mathbf a_t(\theta_j)$. Since
$\boldsymbol{\alpha}\in\mathbb R_+^B$, define
$\overline{\mathbf G}_j
\triangleq\operatorname{Re}\{\mathbf g_j\mathbf g_j^H\}
\succeq\mathbf 0$, leading to
\begin{equation}
\left|\mathbf a_t^H(\theta_j)\mathbf W\boldsymbol{\alpha}\right|^2
=
\boldsymbol{\alpha}^T\overline{\mathbf G}_j\boldsymbol{\alpha}.
\end{equation}
The receive angular coupling is defined as $G^{\mathrm r}_{i,j}
\triangleq
{\left|\mathbf a_r^H(\theta_i)\mathbf a_r(\theta_j)\right|^2}
/{N_r}$.
Write the composite target coefficient as
$\xi_j=\sqrt{\rho_j}e^{\jmath\phi_j}$, where
$\rho_j=|\xi_j|^2$ is the target power coefficient and
$\phi_j$ is the composite propagation and scattering
phase. Since the relative phases of the echoes from different targets are generally unavailable at the transmitter, we model
$\{\phi_j\}_{j\in\mathcal T}$ as mutually independent and
uniformly distributed over $[0,2\pi)$. Hence,
$\mathbb E\{\xi_j\xi_k^{*}\}=0$ for $j\neq k$, and the
phase-averaged aggregate leakage at target $i$'s bin is given by
\begin{equation}
L_i(\boldsymbol{\alpha})
\triangleq
\mathbb E_{\boldsymbol{\phi}}
\{
|\sum_{j\neq i}r_{j\rightarrow i}|^2
\}
=
\sum_{j\neq i}
\mathbb E_{\boldsymbol{\phi}}
\left\{|r_{j\rightarrow i}|^2\right\}.
\end{equation}
Accordingly, the desired focused power and phase-averaged
aggregate leakage can be written as
$S_i(\boldsymbol{\alpha})
=\boldsymbol{\alpha}^{T}\overline{\mathbf R}_i
\boldsymbol{\alpha}$ and
$L_i(\boldsymbol{\alpha})
=\boldsymbol{\alpha}^{T}\overline{\mathbf D}_i
\boldsymbol{\alpha}$, where
$\overline{\mathbf R}_i=\rho_iN_r\overline{\mathbf G}_i$ and
\begin{equation}
\bar{\mathbf D}_i
=
\sum_{j\neq i}
\rho_jG^r_{i,j}
\norm{A(\ell_i-\ell_j,\nu_j-\nu_i)}^2
\bar{\mathbf G}_j.
\label{eq:D_matrix}
\end{equation}
The matrix $\bar{\mathbf R}_i$ captures the desired illumination of target $i$, while $\bar{\mathbf D}_i$ captures the pairwise leakage from the other targets into target $i$'s focused bin. Hence, we define the phase-averaged target-bin visibility of
target $i$ as
\begin{equation}
\Lambda_i(\boldsymbol\alpha)
=
{\boldsymbol\alpha^T\bar{\mathbf R}_i\boldsymbol\alpha}/
({\boldsymbol\alpha^T\bar{\mathbf D}_i\boldsymbol\alpha+\eta}).
\label{eq:visibility}
\end{equation}


We aim to maximize the weakest target visibility. Define 
$\mathbf{q} = [q_1, \ldots, q_B]^T$ with $q_b\in\{0,1\}$ indicating whether beam $b$ is selected, and $K_{s} \leq B$ as the maximum number of selected beams. The problem is then written as
\begin{subequations}
\label{eq:problem}
\begin{align}
\max_{\boldsymbol\alpha,\mathbf q}\quad
&\min_{i\in\mathcal{T}}\; \Lambda_i(\boldsymbol\alpha)\\
\text{s.t.}\quad
&\sum_{b=1}^{B}q_b\leq K_{s},\label{eq:problema}\\
&\alpha_b^2\leq q_bP_t,\qquad b=1,\ldots,B,\label{eq:problemb}\\
&\norm{\boldsymbol\alpha}^2\leq P_t,\label{eq:problemc}\\
&\boldsymbol\alpha\geq\mathbf 0,\label{eq:problemd}\\
&q_b\in\{0,1\},\qquad b=1,\ldots,B.\label{eq:probleme}
\end{align}
\end{subequations}
This is a max-min fractional quadratically constrained program with continuous and binary variables, making it non-convex and difficult to solve in the current form.
\vspace{-1mm}
\section{Optimization Framework}\label{sec:solution}
\vspace{-1mm}
We solve~\eqref{eq:problem} using \ac{mi}-\ac{sca}, with bisection over the visibility level and the non-convex visibility constraints handled by \ac{sca}. The feasibility problem at each SCA step is a mixed-integer convex quadratic program, which can be represented as a \ac{mi}-\ac{socp}.
\vspace{-1mm}
\subsection{Bisection-\ac{mi}-SCA Algorithm}
\vspace{-1mm}
Let $t$ denote a visibility level such that requiring every target to achieve at least $t$ gives
\begin{equation}
\boldsymbol\alpha^T\bar{\mathbf R}_i\boldsymbol\alpha
\geq
 t\boldsymbol\alpha^T\bar{\mathbf D}_i\boldsymbol\alpha+t\eta,
\qquad
 i\in\mathcal T.
\label{eq:non-convex_visibility_constraint}
\end{equation}
For fixed $t$, both quadratic terms are convex in $\boldsymbol\alpha$, but the desired-power term appears on the left-hand side of a lower-bound constraint; hence,~\eqref{eq:non-convex_visibility_constraint} is non-convex. Therefore, we use an \ac{sca} approach over the amplitude variables. Since $\bar{\mathbf R}_i\succeq \mathbf 0$, the desired-power term admits a global affine lower bound around the current point, which can replace the left-hand side conservatively at each iteration.\footnote{Setting $\mathbf X=\boldsymbol\alpha\boldsymbol\alpha^T$ would
yield a rank-one semidefinite formulation, but rank-one recovery and binary variables would result in a poorly scalable \ac{mi} semidefinite program with limited solver support.}


Let $\boldsymbol\alpha^{(r)}$ denote the local point at SCA iteration $r$. A linear local convex approximation of~\eqref{eq:non-convex_visibility_constraint} is written as
\begin{equation}
 t\boldsymbol\alpha^T\bar{\mathbf D}_i\boldsymbol\alpha\!+\!t\eta
\!\leq\!
2(\boldsymbol\alpha^{(r)})^T\bar{\mathbf R}_i\boldsymbol\alpha
\!-\!(\boldsymbol\alpha^{(r)})^T\bar{\mathbf R}_i\boldsymbol\alpha^{(r)},
 i\!\in\!\mathcal T.
\label{eq:sca_visibility_constraint}
\end{equation}
For fixed $t$ and $\boldsymbol\alpha^{(r)}$, the constraint in~\eqref{eq:sca_visibility_constraint} is convex. Thus, the feasibility problem at \ac{sca} iteration $r$ is written as
\begin{subequations}
\label{eq:misocp_feasibility_problem}
\begin{align}
\text{find}\quad
&\boldsymbol\alpha,\mathbf q
\\
\text{s.t.}\quad
& \eqref{eq:problema},\eqref{eq:problemb},\eqref{eq:problemc},\eqref{eq:problemd},\eqref{eq:probleme}, \eqref{eq:sca_visibility_constraint}.
\end{align}
\end{subequations}
This can be handled as a \ac{mi}-\ac{socp} using standard conic solvers, e.g., CVX \cite{cvxref}. 


\begin{algorithm}[t]
\caption{Bisection SCA Beam Selection (B-MI-SCA).}
\label{alg:bisection_misocp}
\begin{algorithmic}[1]
\State \textbf{Input:} $\bar{\mathbf R}_i,\bar{\mathbf D}_i,\forall i\in\mathcal T$, $\eta$, tolerances $\delta$ and $\epsilon$, and $R_{\max}$
\State \textbf{Initialize:} feasible $(\boldsymbol\alpha^{\rm init},\mathbf q^{\rm init})$, $t_{\rm lo}$ and $t_{\rm hi}$,
\State Set $(\boldsymbol\alpha^{\star},\mathbf q^{\star})\gets(\boldsymbol\alpha^{\rm init},\mathbf q^{\rm init})$
\While{$t_{\rm hi}-t_{\rm lo}>\delta$}
\State $t\gets (t_{\rm lo}+t_{\rm hi})/2$, $r = 1$
\State $\boldsymbol\alpha^{(r)}\gets\boldsymbol\alpha^{\rm init}$, $\mathbf q^{(r)}\gets\mathbf q^{\rm init}$, $\Delta\gets\infty$, and $\mathrm{flag}=0$
\While{$\Delta>\epsilon$ and $r<R_{\max}$}
\If{the \eqref{eq:misocp_feasibility_problem} is feasible}
\State Obtain $(\hat{\boldsymbol\alpha},\hat{\mathbf q})$
\State $\Delta\gets \|\hat{\boldsymbol\alpha}-\boldsymbol\alpha^{(r)}\|/\norm{\boldsymbol\alpha^{(r)}}$
\State $\boldsymbol\alpha^{(r)}\gets\hat{\boldsymbol\alpha}$, $\mathbf q^{(r)}\gets\hat{\mathbf q}$, $r \leftarrow r + 1$, and $\mathrm{flag}=1$
\Else{ $\mathrm{flag}=0$, \textbf{break}}
\EndIf
\EndWhile
\If{$\mathrm{flag}=1$}
\State $t_{\rm lo}\gets t$, $(\boldsymbol\alpha^{\star},\mathbf q^{\star})\gets(\boldsymbol\alpha^{(r)},\mathbf q^{(r)})$
\State $(\boldsymbol\alpha^{\rm init},\mathbf q^{\rm init})\gets(\boldsymbol\alpha^{(r)},\mathbf q^{(r)})$
\Else{ $t_{\rm hi}\gets t$}
\EndIf
\EndWhile
\State $\mathcal B^{\star}\gets \{b:q_b^{\star}=1\}$ and $P_b^{\star}\gets (\alpha_b^{\star})^2$
\State \Return $\mathcal B^{\star}$, $\boldsymbol\alpha^{\star}$, and $t_{\rm lo}$
\end{algorithmic}
\end{algorithm}

The bisection-\ac{mi}-SCA procedure is summarized in Algorithm~\ref{alg:bisection_misocp}. For each trial visibility level $t$, the inner loop solves a local approximation of the original
fixed-$t$ constraints. Whenever \eqref{eq:misocp_feasibility_problem} is feasible, the resulting solution is also feasible for the original fixed-$t$ problem, and the achieved lower bound can be increased. Conversely, infeasibility of the local inner approximation does not certify global infeasibility of the
original fixed-$t$ problem. Therefore, Algorithm~\ref{alg:bisection_misocp} returns a feasible beam-selection and power-allocation solution together with an achieved lower bound on the optimal worst-target visibility, but it does not provide a global guarantee, while its performance can depend on the initialization \cite{Boyd2004}.


\vspace{-1mm}
\subsection{Time-Sharing Perspective}
\vspace{-1mm}


In the time-sharing strategy, the selected beams are activated in separate full-grid sensing intervals rather than transmitted simultaneously. This provides a standalone alternative with the sensing intervals distributed across the selected codebook beams in time. For a single active beam $b$, define the transmit and receive
angular gains as
$G^{\mathrm t}_{j,b}
=|\mathbf a_t^H(\theta_j)\mathbf w_b|^2$
and $G^{\mathrm r}_{i,j}$, respectively.
The desired focused power of target $i$ under beam $b$ is written as $S_{i,b}=P_t\rho_i N_r G^t_{i,b}$,
while the noncoherent leakage power at target $i$'s focused bin under beam $b$ is given by
\begin{equation}
L_{i,b}
=
P_t\sum_{j\neq i}
\rho_j
G^r_{i,j}
G^t_{j,b}
\norm{A(\ell_i-\ell_j,\nu_j-\nu_i)}^2.
\label{eq:L}
\end{equation}
By defining the time-sharing vector $\boldsymbol{\tau}=[\tau_1,\ldots,\tau_B]$, with $\sum_{b=1}^{B}\tau_b=1$, the visibility of target $i$ is defined as
\begin{equation}
\Lambda_i^{\rm ts}(\boldsymbol{\tau})=
\frac{\sum_{b=1}^{B}\tau_b S_{i,b}}
{\sum_{b=1}^{B}\tau_b L_{i,b}+\eta}.
\label{eq:ts_visibility}
\end{equation}
Thus, the time-sharing problem is written as
\begin{subequations}
\label{eq:ts_problem}
\begin{align}
\max_{\boldsymbol{\tau},\mathbf q}\quad
&\min_{i\in\mathcal{T}}\; \Lambda_i^{\rm ts}(\boldsymbol{\tau})\\
\text{s.t.}\quad
&0\leq \tau_b\leq q_b,\qquad b=1,\ldots,B,\label{eq:ts_problemb}\\
&\sum_{b=1}^{B}\tau_b=1,\label{eq:ts_problemc}\\
&\eqref{eq:problema},\eqref{eq:probleme}.
\end{align}
\end{subequations}
Here, the visibility constraints become linear once $t$ is fixed, leading to $\Lambda_i^{\rm ts}(\boldsymbol{\tau})\geq t$ being equivalent to $\sum_{b=1}^{B}
\tau_b
\left(
S_{i,b}-tL_{i,b}
\right)
\geq
 t\eta$. Thus, for fixed $t$, the feasibility problem becomes an \ac{mi} \ac{lp}. The bisection-\ac{mi}\ac{lp} procedure for the time-sharing system is summarized in Algorithm~\ref{alg:bisection_milp}. 


The time-sharing solution can also provide a structured initialization for Algorithm~\ref{alg:bisection_misocp}. Specifically, the beam support is obtained from B-MILP-TS, and the time-sharing fractions are mapped to simultaneous amplitudes as $\alpha_b=\sqrt{P_t \tau_b^\star}, b\in\mathcal{B}^\star$, and $\alpha_b=0$ otherwise. This mapping satisfies the power constraint~\eqref{eq:problemc} because $\sum_b \tau_b^\star=1$. Although the time-sharing and simultaneous-transmission problems are not equivalent, the mapped point provides a structured initialization point.

\begin{algorithm}[t]
\caption{Bisection-\ac{mi}\ac{lp} Time-Sharing Beam Selection (B-\ac{mi}\ac{lp}-TS).}
\label{alg:bisection_milp}
\begin{algorithmic}[1]
\State \textbf{Initialize:} $S_{i,b}, L_{i,b}, \forall\{i,b\}$, noise floor $\eta$, tolerance $\delta$, $t_{\rm lo}$, $t_{\rm hi}^{\rm ts}$, $\boldsymbol{\tau}^{\star}\gets \varnothing$, and $\mathbf q^{\star}\gets \varnothing$
\While{$t_{\rm hi}^{\rm ts}-t_{\rm lo}>\delta$}
\State $t\gets (t_{\rm lo}+t_{\rm hi}^{\rm ts})/2$
\State Solve the \ac{mi}\ac{lp} feasibility problem
\If{the \ac{mi}\ac{lp} is feasible}
\State Store the feasible solution as $(\boldsymbol{\tau}^{\star},\mathbf q^{\star})$, $t_{\rm lo}\gets t$
\Else{ $t_{\rm hi}^{\rm ts}\gets t$}
\EndIf
\EndWhile
\State $\mathcal{B}^{\star}\gets \{b:q_b^{\star}=1\}$
\State \Return $\mathcal{B}^{\star}$, $\boldsymbol{\tau}^{\star}$, and $t_{\rm lo}$
\end{algorithmic}
\end{algorithm}


\subsection{Convergence and Complexity}


The approximation in~\eqref{eq:sca_visibility_constraint} is tight at the
current point and constitutes a global lower bound of the
desired term. Hence, every feasible solution of the
MI-SOCP satisfies the original fixed-$t$ visibility constraints.
The bisection and SCA iteration limits guarantee
termination, and the returned visibility level is a feasible
achieved lower bound on the optimum.

Let $N_{\rm b}$ denote the number of bisection iterations and $N_{\rm sca}$ the maximum number of \ac{sca} iterations used in each fixed-$t$ feasibility test, then,
$N_{\rm b}
=
\left\lceil
\log_2
\left(
{(t_{\rm hi}-t_{\rm lo})}/{\delta}
\right)
\right\rceil$. Thus, Algorithm~\ref{alg:bisection_misocp} requires at most $N_{\rm b}N_{\rm sca}$ \ac{mi}-\ac{socp} feasibility tests \cite{Boyd2004}. Each \ac{mi}-\ac{socp} contains $B$ continuous amplitude variables and $B$ binary beam-selection variables, leading to the worst-case complexity exponential in $B$. 

For the time-sharing formulation, the fixed-$t$ feasibility problem is an \ac{mi}\ac{lp} with $B$ continuous time-sharing and binary selection variables. Since the feasibility set is monotone in $t$, bisection returns the global optimum of~\eqref{eq:ts_problem} within tolerance $\delta$. Its complexity is determined by $N_{\rm b}^{\rm ts}$ feasibility tests, obtained over $[0,t_{\rm hi}^{\rm ts}]$ as $N_b^{\rm ts}
=
\left\lceil
\log_2\!\left({t_{\rm hi}^{\rm ts}}/{\delta}\right)
\right\rceil$.

\section{Numerical Results}\label{sec:results}


We consider $f_c=28\,\mathrm{GHz}$ with
$N_t=N_r=32$ antennas and an $N_t$-beam
\ac{dft} codebook. The \ac{ofdm} grid has $N=64$ subcarriers
with $\Delta f=120\,\mathrm{kHz}$ and $M=32$ symbols. Neglecting
the \ac{cp} overhead gives $T_{\rm sym}=1/\Delta f\simeq
8.33\,\mu\mathrm{s}$, delay resolution
$\delta_\ell=1/(N\Delta f)\simeq130\,\mathrm{ns}$, and Doppler
resolution $\delta_\nu=1/(MT_{\rm sym})\simeq3.7\,\mathrm{kHz}$
\cite{3gpp_tr_22870_v2000}. The transmit \ac{snr} is set to
$10\,\mathrm{dB}$ with unit noise variance, yielding $P_t=10$.


Each scene contains $I=4$ targets arranged as two weak/strong pairs. The received power coefficients are set as $\rho_w=10^{c_w/10}$ and $\rho_s=10^{c_s/10}$, where $c_w\sim\mathcal{U}[-11,-9]~\mathrm{dB}$ and $c_s\sim\mathcal{U}[9,11]~\mathrm{dB}$, yielding an average power contrast of $20~\mathrm{dB}$ per pair. This models a scenario in which a weak target (e.g., a pedestrian) and a nearby strong target (e.g., a vehicle) differ significantly in their composite scattering return. Note that $\Lambda_i(\boldsymbol{\alpha})$ depends on the inter-target power ratios $\rho_j/\rho_i$ and the noise-normalized signal level $\rho_i N_r P_t / \eta$ rather than the absolute values of $\rho_i$ individually. Hence, scaling all $\rho_i$ values by a common factor is equivalent to a shift in the \ac{snr}. For each pair, $\theta_w\sim\mathcal{U}[-55^\circ,55^\circ]$, while the strong target is placed at $
\sin\theta_s
=
\sin\theta_w
\pm{2\delta_\theta}/{N_t},$ so that $\delta_\theta$ measures angular separation in DFT-beamwidth units in sine space. The weak target \ac{dd} coordinates are drawn as $\ell_w\sim\mathcal{U}[8,N-8]$ and $\nu_w\sim\mathcal{U}[-M/4,M/4]$, and the strong target is offset by $\delta_{\rm DD}$ bins in both delay and Doppler. Unless otherwise stated, $\delta_\theta=0.5$, $\delta_{\rm DD}=0.5$, and $K_s=4$. All results are averaged over $100$ independent random scenes.



We consider the following baselines:\\
\textbf{LS} minimizes the aggregate leakage power without a selected-beam limit by using $\mathbf{X}=\boldsymbol{\alpha}\boldsymbol{\alpha}^{\top}$ and solving
\[
\min_{\mathbf{X}\succeq 0}
\operatorname{Tr}\!\left(\bar{\mathbf{D}}_{\rm glob}\mathbf{X}\right)
\quad
\text{s.t.}
\operatorname{Tr}(\mathbf{X})\leq P_t,
\operatorname{Tr}(\bar{\mathbf{R}}_i\mathbf{X})\geq \eta,\ \forall i,
\]
where $\bar{\mathbf{D}}_{\rm glob}=\sum_i\bar{\mathbf{D}}_i$ and rank-one recovery is performed by Gaussian randomization. This baseline represents a global sidelobe-suppression-based design \cite{moveble_isl,SLP_sidelobe} since it minimizes aggregate leakage rather than the worst-target visibility.\\
\textbf{Top-$K_s$} selects the $K_s$ beams with the largest aggregate desired illumination score
\[
c_b=\sum_i [\bar{\mathbf{R}}_i]_{b,b}
=\sum_i \rho_i N_r |\mathbf{a}_t^{\mathsf H}(\theta_i)\mathbf{w}_b|^2 .
\]
The selected beams are assigned equal amplitudes $\alpha_b=\sqrt{P_t/K_s}$. This provides an illumination-oriented baseline, motivated by conventional sensing beamforming designs that optimize beampattern gains \cite{opt_BF_ISAC}, ignoring inter-target leakage.\\
\textbf{Random} selects $K_s$ beams uniformly at random and assigns equal amplitudes $\alpha_b=\sqrt{P_t/K_s}$ to the selected beams.

\begin{figure}[t]
  \centering  \includegraphics[width=0.9\columnwidth]{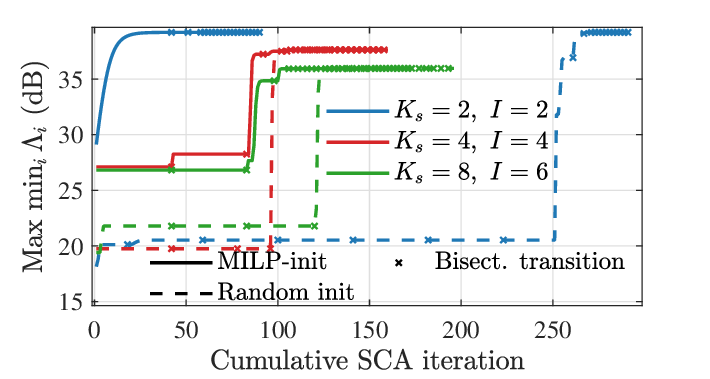}
  \vspace{-1mm}
\caption{Convergence behavior of the proposed B-\ac{mi}-SCA algorithm versus cumulative SCA iterations.}
\label{fig:conv}  \label{fig:conv}
\vspace{-5mm}
\end{figure}

Fig.~\ref{fig:conv} shows the convergence behavior of B-\ac{mi}-\ac{sca} for different $K_s$ and $I$ under \ac{mi}\ac{lp}-based and random initializations. As expected, the \ac{mi}\ac{lp}-based initialization provides a stronger feasible starting point and reaches high visibility levels with fewer \ac{sca} iterations. Random initialization may stay at low visibility levels for longer, especially for larger $K_s$ and $I$, but it eventually reaches the same solution in these cases. This empirically indicates limited sensitivity to initialization in the considered settings, and the time-sharing solution provides a more efficient structured starting point.

\begin{figure}[t]
  \centering  \includegraphics[width=0.9\columnwidth]{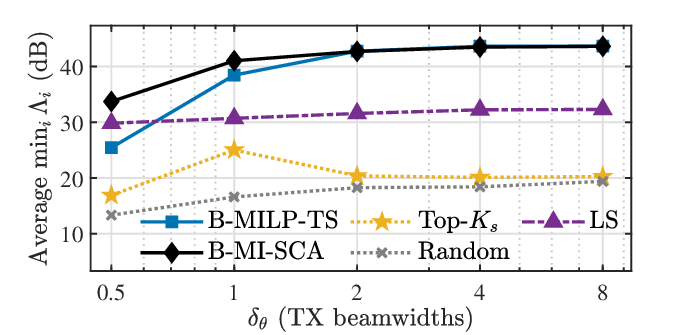}
  \vspace{-1mm}
\caption{Average worst-target visibility versus $\delta_\theta$.}
  \label{fig:ang}
  \vspace{-5mm}
\end{figure}

Fig.~\ref{fig:ang} shows the impact of angular separation on the average worst-target visibility. When $\delta_\theta$ is small, B-\ac{mi}-\ac{sca} has clear gain over all baselines, including B-MILP-TS, because simultaneous probing can exploit spatial degrees of freedom through coherent beam superposition, whereas time sharing only allocates sensing fractions across beams. As $\delta_\theta$ increases, the targets become more separable by the finite codebook, the angular coupling weakens, and the gap between simultaneous probing and time sharing decreases. This shows that time sharing becomes nearly sufficient when the targets are well separated spatially, while simultaneous beam selection is most useful in angularly dense scenes.

\begin{figure}[t]
  \centering  \includegraphics[width=0.9\columnwidth]{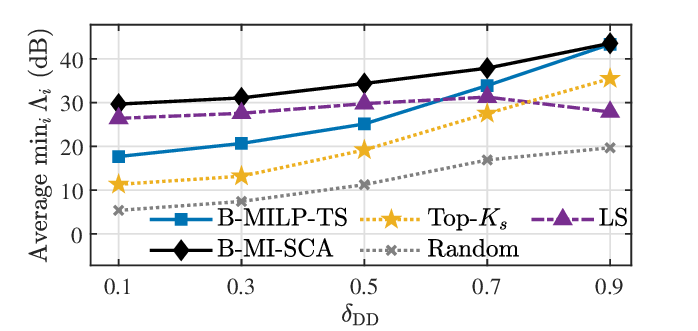}
  \vspace{-1mm}
\caption{Average worst-target visibility versus $\delta_{\rm DD}$.}
\label{fig:dd}
\vspace{-5mm}
\end{figure}

Fig.~\ref{fig:dd} shows the effect of \ac{dd} separation. For small $\delta_{\rm DD}$, B-\ac{mi}-SCA clearly outperforms B-\ac{mi}\ac{lp}-TS because simultaneous probing can jointly shape the illumination of desired and interfering targets in proximity. As $\delta_{\rm DD}$ increases, the ambiguity coupling decreases, and all methods improve. Consequently, the gap between B-\ac{mi}-SCA and B-\ac{mi}\ac{lp}-TS becomes smaller, since time sharing is sufficient when leakage is weak. Top-$K_s$ remains unreliable because it ignores the leakage, while LS is clearly below B-\ac{mi}-SCA and has a different behavior because it minimizes aggregate leakage.

\begin{figure}[t]
  \centering  \includegraphics[width=0.9\columnwidth]{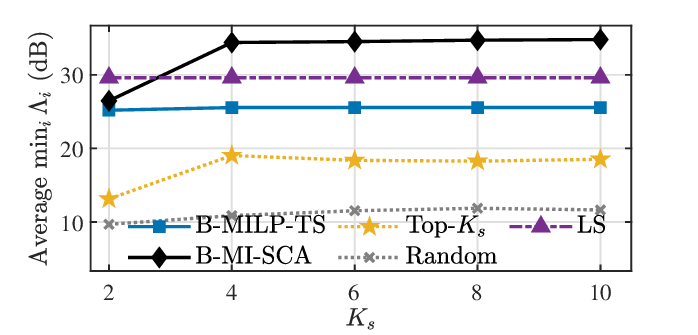}
  \vspace{-1mm}
\caption{Average worst-target visibility versus $K_s$.}
  \label{fig:sweep_ksel}
  \vspace{-5mm}
\end{figure}

Fig.~\ref{fig:sweep_ksel} shows the effect of $K_s$. For $K_s=2$, B-\ac{mi}-SCA performs below LS because LS is not subject to the beam-selection limit and can exploit all codebook beams, which also explains its constant pattern across $K_s$. The proposed B-\ac{mi}-SCA method improves when $K_s$ increases from $2$ to $4$, since the additional beams provide enough spatial degrees of freedom to balance the illumination of weak targets while reducing leakage from nearby strong targets. Beyond $K_s=4$, the gain saturates, indicating that the most useful beams have already been selected for the considered four-target scenes. B-\ac{mi}\ac{lp}-TS is almost insensitive to $K_s$ because the time-sharing solution allocates its resources to the most useful beams and cannot exploit coherent simultaneous beam superposition.

\begin{figure}[t]
  \centering\includegraphics[width=0.9\columnwidth]{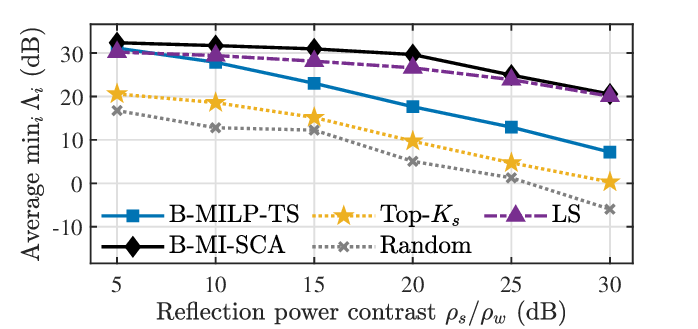}
  \vspace{-1mm}
  \caption{Average worst-target visibility versus reflection power contrast.}
  \label{fig:sweep_contrast}
  \vspace{-5mm}
\end{figure}

Fig.~\ref{fig:sweep_contrast} shows the impact of weak/strong reflection-power contrast. As $\rho_s/\rho_w$ increases, leakage from the strong target increasingly masks the weak target, and the visibility decreases for all methods. The proposed B-MI-SCA remains the most robust because it explicitly balances weak-target illumination against leakage from nearby strong targets. The gain is most evident in the medium-contrast regime, where masking is already significant but can still be mitigated by simultaneous beam superposition. At very low contrast, the masking effect is weaker, while at very high contrast, the leakage becomes dominant and all methods degrade.

\begin{figure}[t]
  \centering  \includegraphics[width=0.9\columnwidth]{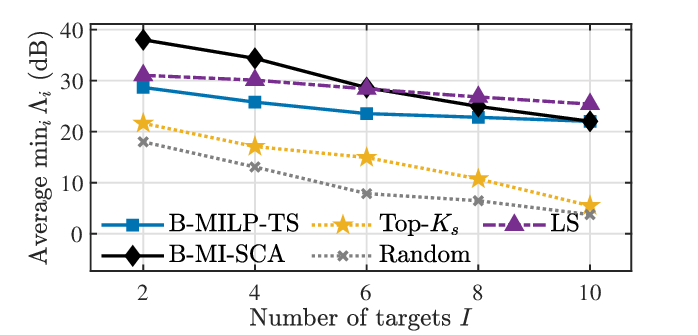}
  \vspace{-1mm}
\caption{Average worst-target visibility versus $I$.}
\label{fig:num_targets}
\vspace{-5mm}
\end{figure}

Fig.~\ref{fig:num_targets} shows the impact of $I$. Because increasing $I$ makes the problem more constrained given a fixed $K_s$, as the same limited set of beams must preserve the visibility of more target bins while also controlling more pairwise leakage terms, the average worst-target visibility decreases for all methods. The proposed B-\ac{mi}-SCA method provides the highest visibility for sparse and moderately dense scenes, where the simultaneous sparse beam superposition can effectively balance target illumination and leakage suppression. However, its advantage decreases with $I$ because the fixed beam-selection limit becomes more restrictive relative to the number of target directions and leakage relationships that must be jointly managed. The B-\ac{mi}\ac{lp}-TS benchmark exhibits a similar decreasing trend but remains below B-\ac{mi}-SCA for small-moderate $I$. The LS baseline is less sensitive at large $I$, as it is not subject to the same $K_s$ and thus represents a structurally different leakage-suppression reference.

\section{Conclusions}\label{sec:conclusion}

This paper proposed a target-bin visibility-oriented codebook beam-selection framework for multi-target \ac{mimo}-\ac{ofdm} sensing. The proposed formulation directly protects each hypothesized target at its own \ac{dd} bin by accounting for the desired focused power and the pairwise leakage from other targets by solving a max-min beam-selection problem using a bisection-\ac{mi}-SCA algorithm. We also presented a time-sharing formulation as an optimally solvable baseline and structured initialization for SCA. The numerical results showed that the proposed B-MI-SCA provides clear gains in close angular or \ac{dd} target separation regimes.

\bibliographystyle{IEEEtran}
\bibliography{ref_abbv}
\end{document}

%% file: acronyms.tex
\DeclareAcronym{isac}{
  short = ISAC,
  long  = integrated sensing and communications
}
\DeclareAcronym{rf}{
  short = RF,
  long  = radio frequency
}
\DeclareAcronym{bs}{
  short = BS,
  long  = base station
}
\DeclareAcronym{genai}{
  short = GenAI,
  long  = generative AI
}
\DeclareAcronym{lidar}{
  short = LiDAR,
  long  = light detection and ranging
}
\DeclareAcronym{6g}{
  short = 6G,
  long  = sixth generation
}
\DeclareAcronym{5g}{
  short = 5G,
  long  = Fifth generation
}

\DeclareAcronym{ai}{
  short = AI,
  long  = artificial intelligence
}
\DeclareAcronym{3gpp}{
  short = 3GPP,
  long  = 3rd generation partnership project
}
\DeclareAcronym{slam}{
  short = SLAM,
  long  = simultaneous localization and mapping
}
\DeclareAcronym{csi}{
  short = CSI,
  long  = channel state information
}
\DeclareAcronym{ri}{
  short = R\&I,
  long  = research and innovation
}

\DeclareAcronym{jepa}{
  short = JEPA,
  long  = joint-embedding predictive architecture
}

\DeclareAcronym{mimo}{
  short = MIMO,
  long  = multiple-input multiple-output
}

\DeclareAcronym{ofdm}{
  short = OFDM,
  long  = orthogonal frequency-division multiplexing
}

\DeclareAcronym{ula}{
  short = ULA,
  long  = uniform linear array
}

\DeclareAcronym{cp}{
  short = CP,
  long  = cyclic prefix
}

\DeclareAcronym{rcs}{
  short = RCS,
  long  = radar cross-section
}

\DeclareAcronym{milp}{
  short = MILP,
  long  = mixed-integer linear programming
}

\DeclareAcronym{mi}{
  short = MI,
  long  = mixed-integer
}

\DeclareAcronym{lp}{
  short = LP,
  long  = linear programming
}

\DeclareAcronym{nr}{
  short = NR,
  long  = new radio
}

\DeclareAcronym{isl}{
  short = ISL,
  long  = integrated sidelobe level
}

\DeclareAcronym{snr}{
  short = SNR,
  long  = signal-to-noise ratio
}

\DeclareAcronym{cdf}{
  short = CDF,
  long  = cumulative distribution function
}

\DeclareAcronym{dd}{
  short = DD,
  long  = delay-Doppler
}

\DeclareAcronym{sca}{
  short = SCA,
  long  = successive convex approximation
}

\DeclareAcronym{socp}{
  short = SOCP,
  long  = second-order conic program
}

\DeclareAcronym{dft}{
  short = DFT,
  long  = Discrete Fourier Transform
}